\documentclass[%
notitlepage,
reprint,
superscriptaddress,
amsmath,amssymb,
aps,
]{revtex4-1}

\usepackage{lipsum}
\usepackage{xcolor}
\usepackage{graphicx}
\usepackage{dcolumn}
\usepackage{bm}

\usepackage{epstopdf}

\newcommand{\bra}[1]{\langle #1 \vert}
\newcommand{\ket}[1]{\vert #1 \rangle}

\begin{document}
	\title{Universal dynamical scaling laws in three-state quantum walks}
	\author{P. R. N. Falc\~ao}
	\email{pedro.falcao@fis.ufal.br}
	\affiliation{%
		Instituto de F\'{i}sica, Universidade Federal de Alagoas, 57072-900 Macei\'{o}, AL, Brazil
	}%
	\author{A. R. C. Buarque}
	\affiliation{%
		Instituto de F\'{i}sica, Universidade Federal de Alagoas, 57072-900 Macei\'{o}, AL, Brazil
	}%
	\author{W. S. Dias}
	\affiliation{%
		Instituto de F\'{i}sica, Universidade Federal de Alagoas, 57072-900 Macei\'{o}, AL, Brazil
	}%
	\author{G. M. A. Almeida}
	\affiliation{%
		Instituto de F\'{i}sica, Universidade Federal de Alagoas, 57072-900 Macei\'{o}, AL, Brazil
	}%
	\author{M. L. Lyra}
	\affiliation{%
		Instituto de F\'{i}sica, Universidade Federal de Alagoas, 57072-900 Macei\'{o}, AL, Brazil
	}%
	\begin{abstract}
		We perform a finite-time scaling analysis over the detrapping point of a three-state quantum walk on the line. The coin operator is parameterized by $\rho$ that controls the wavepacket spreading velocity. The input state prepared at the origin is set as
		symmetric linear combination of two eigenstates of the coin operator with a characteristic mixing angle $\theta$, one of them being the component that results in full spreading occurring at $\theta_c(\rho)$ for which no fraction of the wavepacket remains trapped near the initial position. 
		We show that relevant quantities such as the survival probability and the participation ratio assume single parameter scaling forms at the vicinity of the detrapping angle $\theta_c$. In particular, we show that the participation ratio grows linearly in time with a logarithmic correction, 
		thus shedding light on previous reports of sublinear behavior.

	\end{abstract}
	\maketitle
	
	\section{Introduction} 
	For many decades, classical random walks have been essential for the development of algorithms, as well for describing a wide range of physical phenomena. 
	The story is no different for their quantum counterpart, namely quantum walks, since put forward by Aharonov \textit{et al.} in \cite{aharonov1993}. They have been crucial to the progress of quantum computing, including the design of quantum algorithms \cite{shenvi2003,childs2004,childs2014,wong2015,wong2018} for search problems and state transfer \cite{zhan2014, stefanak2016,kurzynsky2011}, to name a few. 
	One of the many advantages is the speedup gained from the fact that the quantum walker spreads out ballistically due to interference between its wavefunction amplitudes \cite{knight2003}.
	Moreover, it was shown both continuous and discrete versions make for universal quantum computation \cite{childs2009,lovett2010}. 
	
	In the discrete-time quantum walk, particularly, the walker evolves through a predefined sequence of gates. The first gate is usually the coin operator that superposes its internal degrees of freedom, followed by the translation operator that moves it (say, left or right) depending on its internal state. This goes on and on until the protocol reaches its goal \cite{kempe2003,venegas2012}. 
	The key point here is that there are endless forms to set those rules and thus different routes to travel through the Hilbert space, what drives progress in assessing hard computational problems \cite{vakulchyk2019}.  
	From that point of view, discrete-time quantum walks are also a powerful tool for quantum simulation of complex phenomena such as quantum phase transitions \cite{chandrashekar2008,wang2019,xwang2018},
	nonlinear dynamics \cite{navarrete2007,buarque2020,mendonca2020,buarque2021}, topological phases of matter \cite{kitagawa2010,asboth2012,rakovszky2015}, 
	localization \cite{obuse2011,mendes2019,buarque2019,queiros2020}, and many others.

	A remarkable feature is that in most of the cases pointed out above, the quantum walk setups giving rise to that kind of phenomena are simple two-state models on a line. We may get even more of it by adding, for instance, extra internal degrees of freedom. 
	In this context, the three-state Grover walk \cite{inui2005} has attracted some attention due to its many ubiquitous properties, one of them being an intrinsic localization taking place at the input site, something that just does not occur to the classical random walker. 
	In this Grover walk, the walker can move
	to the left, right, and stay at the same position.
	The additional internal degree of freedom 
	is responsible for the generation of a constant eigenvalue in Fourier space which leads to a nonvanishing probability at the initial input position.
	Thus, the probability to find the walker at this position will saturate to a constant value in the long time regime.
	\cite{inui2005,falkner2014, konno2014}.
	%
	%
	Two generalizations of the Grover walk was obtained in \cite{stefanak2012} by
	working out families of three-state coin operators that maintain localization. The same authors investigated the quantum walk dynamics in  detail in \cite{stefanak2014}, discussing the role played by each eigenvector on the dynamics via an analytical approach.
	Moreover, it was shown in \cite{wong2017} that a class of lackadaisical quantum walks (where each site features an integer number of self-loops) is equivalent to one of those continuous deformations of the three-state Grover coin proposed in \cite{stefanak2012}.
	Due to its faster ballistic dispersion, this kind of quantum walk can boost search algorithms in general graphs, as demonstrated in \cite{wong2015,wong2018}. 
	More recently, an implementation of three-state quantum walks in a quantum circuit model was proposed in \cite{saha2021}.   
	
	In this work, we 
	seek to unveil some similarities 
	between a general form of the three-state Grover walk with other classes of discrete-time quantum walks by looking for dynamical scaling laws
	at the vicinity of
	the threshold where localization is suppressed. To do so, we define a set
	of effective two-component initial states whose relative phase is defined by a mixing angle $\theta$.
	At the threshold $\theta=\theta_c$ the dynamics resembles that of the standard two-state Hadamard quantum walk for it
	evolves ballistically with no localized component. We show that the survival probability and the participation ratio satisfies universal dynamical scaling laws in the vicinity of $\theta_c$ for any value of the coin deformation parameter. We also obtain an analytical expression for the participation ratio featuring a logarithm correction. This has never been reported in the quantum walk literature and actually explains 
	the emergence of the effective sublinear time evolution seen in \cite{omanakuttan2018}.
	
	\section{Model}
	We consider a quantum walker propagating through an infinite one-dimensional lattice. The Hilbert space of the system is made up of two parts, namely $H = H_p \otimes H_c$, where $H_p$ accounts for the position space $\left\lbrace{|x\rangle} \right\rbrace$, with $x \in \mathbb{Z}$, and $H_c$ is a three-dimensional coin space, spanned by $\left\lbrace {|L \rangle = (1,0,0)^{T}, |S \rangle = (0,1,0)^{T}, |R \rangle = (0,0,1)^{T}}\right\rbrace$. 
	%
	The walker evolves in discrete time steps via $\ket{\psi(t+1)}=\hat{U}\ket{\psi(t)}$, where $\hat{U}=\hat{S}\cdot[\hat{C}\otimes\mathbb{I}]$ is a unitary operator, with
	$\mathbb{I}$ being the identity operator acting on $H_p$.  
	At each step, following the coin operator $\hat{C}$ responsible for mixing the internal degrees of freedom, the operator
	\begin{eqnarray}
		\hat{S} = \sum_{x =-\infty}^{\infty}&&\bigg[ |x-1\rangle \langle x| \otimes |L \rangle \langle L | + |x\rangle \langle x| \otimes |S \rangle \langle S| \nonumber\\ 
		&& + |x+1\rangle \langle x| \otimes |R \rangle \langle R \mid \bigg]
	\end{eqnarray}
	shifts each position state to the left, right, or none, according to its related internal state. The probability to find the walker at position $x$ at a given step $t$ is 
	given by $P_{x}(t) = \sum_{c}\vert(\bra{x}\otimes \bra{c}) \ket{\psi(t)}  \vert^2$, with $c \in \{L,S,R\}$. 

	The three-state coin operator we consider here is a one-parameter class of the Grover diffusion coin that preserves localization (see \cite{stefanak2012} for details), namely
	\begin{equation}
		\hat{C}(\rho) = 
		\begin{pmatrix}  
			-\rho^{2} & \rho\sqrt{2 - 2\rho^{2}} & 1 - \rho^{2} \\ 
			\rho\sqrt{2 - 2\rho^{2}} & 2\rho^{2} -1 & \rho\sqrt{2 - 2\rho^{2}} \\
			1 - \rho^{2} & \rho\sqrt{2 - 2\rho^{2}} & -\rho^{2}
		\end{pmatrix},
		\label{Eq.2}
	\end{equation}
	where the coin parameter $\rho \in (0,1)$ determines the speed of ballistic dispersion. That means one will find the peaks at $\pm \rho t$ after $t$ steps.  
	Note that $\rho=0$ yields trivial dynamics, as $\hat{C}$ is reduced to the permutation matrix, meaning that
	the walker will get stuck between the origin and its nearest neighbors. On the other hand, if $\rho=1$ the
	coin operator gets the diagonal form and no shuffling will occur during evolution. The standard, unweighted Grover coin \cite{inui2005} is recovered when $\rho=1/\sqrt{3}$.
	
	At this point, we shall work out the eigenvectors of the coin operator defined above, as expressing the walker input state in such basis will simplify the analysis considerably.
	Straightforward algebra leads to 
	\begin{align}
		\ket{\sigma^+} &= \sqrt{\frac{1-\rho^2}{2}}\ket{L} + \rho\ket{S} + \sqrt{\frac{1-\rho^2}{2}}\ket{R},\\
		\ket{\sigma_1^-} &= \frac{\rho}{\sqrt{2}}\ket{L} - \sqrt{1-\rho^2} \ket{S} + \frac{\rho}{\sqrt{2}}\ket{R},\\
		\ket{\sigma_2^-} &= \frac{1}{\sqrt{2}}(\ket{L} - \ket{R}),
	\end{align}
	corresponding to eigenvalues 1,-1,-1, respectively. An initial state prepared at $x=0$ can thus be conveniently written as $\ket{\psi(t=0)} = \ket{0}\otimes \ket{\psi_{c}}$, with
	\begin{equation}\label{psi_c}
		\ket{\psi_{c}}=\alpha\ket{\sigma^{+}}+\beta\ket{\sigma_{1}^{-}}+\gamma\ket{\sigma_{2}^{-}}.
	\end{equation}
	Each of these parts plays very specific roles in the dynamics, 
	as shown in \cite{stefanak2014} in great detail.
	{In particular, the initial state $\ket{\sigma_1^-}$ alone does not lead to
		localization at the origin. Rather, its propagation resembles that of the two-state Hadamard quantum walk. 
		Outcomes for each coin state can be assessed analytically via Fourier analysis and weak-limit theorems \cite{grimmet2004}. One can then approximate the spatial probability distribution in the long-time limit in terms of the group velocity density $\omega(\nu)$. This was worked out 
		first for the special case featuring $\rho=1/\sqrt{3}$ \cite{falkner2014} and later for arbitrary $\rho$ using the  Riemann-Lebesgue lemma \cite{machida2015}.
		For $\ket{\psi_{c}}=\ket{\sigma_1^-}$, the group velocity density reads \cite{stefanak2014}
		\begin{equation}
			\omega(\nu) = \frac{\sqrt{1 - \rho^2}}{\pi(1-\nu^2)\sqrt{\rho^2 - \nu^2}} ,
			\label{Eq.3}
		\end{equation}
		where $\nu = x/t$. The probability distribution for finite $t$ follows directly as $P_{x}(t) = \omega(\nu)/t$. This readily indicates that the probability to find the particle at the origin goes to zero with time.}
	
	\section{Results}
	
	Our primary goal here is to look after dynamical scaling laws at the vicinity of the point where localization starts to take place. It means that 
	the initial coin state $\ket{\psi_{c}}$ must overlap with at least another eigenstate besides $\ket{\sigma_1^-}$. 
	If we set at the origin ($x=0$) 
	a symmetric initial state of the form
	\begin{equation}
		\ket{\psi(t=0)}=\ket{0}\otimes \left( \cos \theta \ket{S} + \sin \theta \ket{\phi} \right),
	\end{equation}
	with $\ket{\phi} = (\ket{L}+\ket{R})/\sqrt{2}$ and $\theta \in [0, \pi]$, the coefficients of Eq. (\ref{psi_c}) become $\alpha = (\rho \cos \theta + \sqrt{1 - \rho^2}\sin \theta)$, $\beta = (\rho \sin \theta - \sqrt{1 - \rho^2}\cos \theta)$, and $\gamma=0$. We thus define a characteristic angle $\theta_c = \cos^{-1}(-\sqrt{1 - \rho^2})$, which is the angle for which $\alpha=0$ and $|\psi_0 \rangle = |\sigma_1^{-} \rangle$. By definition, $\theta_c \in [\pi/2, \pi]$. 
	
\begin{figure}[t!]
		\centering
		\includegraphics[width = 0.45 \textwidth]{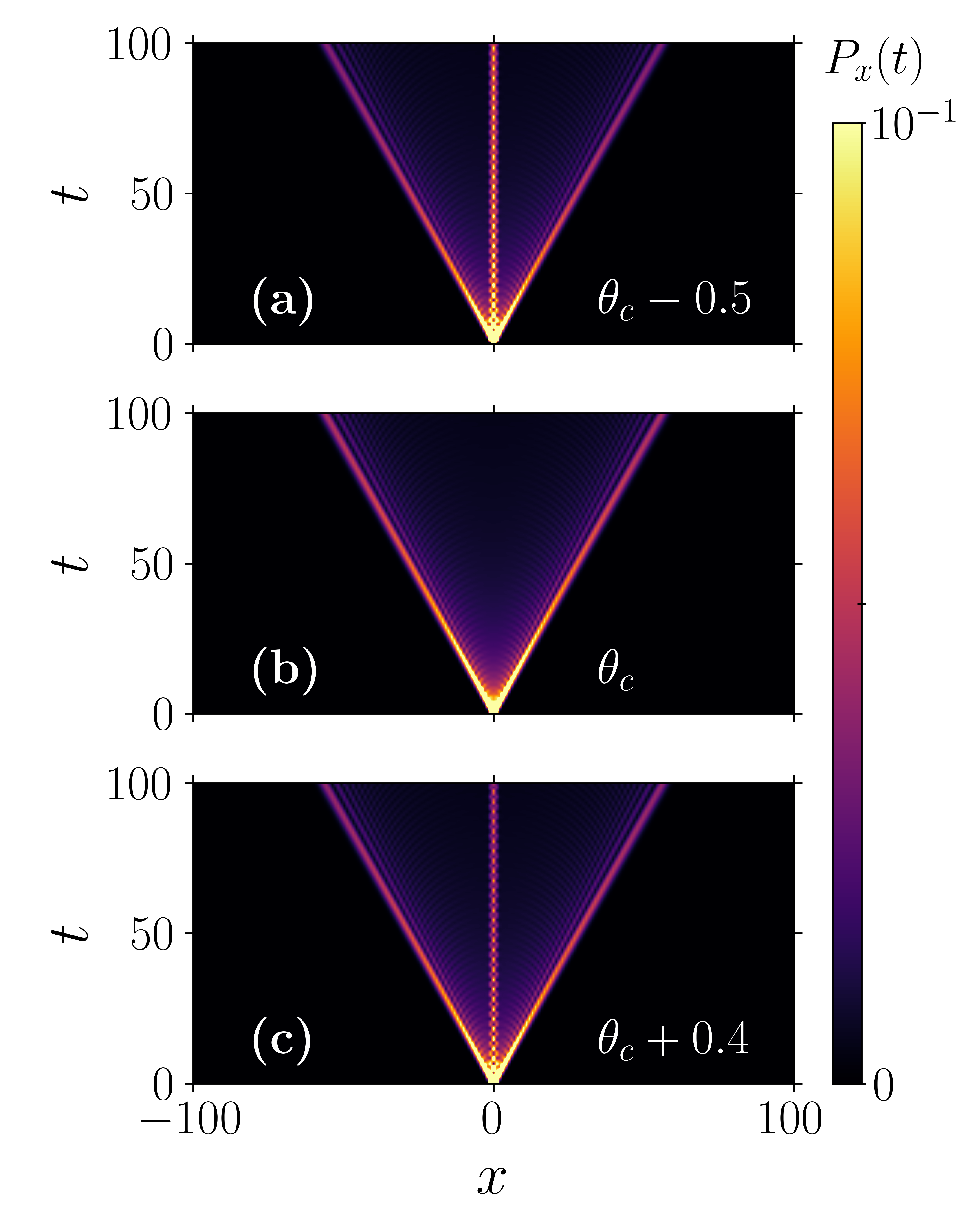}
		\caption{Time evolution of occupation probability $P_{x}$ for the three-state Grover Walk ($\rho = 1/\sqrt{3}$) over $100$ time-steps for distinct values of $\theta$. In general, part of the wavepacket remains trapped at the origin, except at $\theta_c$ which is when the initial coin state corresponds to $|\sigma_1^{-}\rangle$.}
		\label{Fig.1}		
	\end{figure}

	We are now ready to track the dynamics of such input walker state featuring a mixing angle $\theta$ between 
	eigenstates that lead to localized and delocalized behavior \cite{stefanak2014}. We set open boundary conditions and sizes $N$ large enough so as to avoid the wavefunction from reaching the edges.  
	Figure \ref{Fig.1} shows the probability $P_{x}(t)$ evolution in space and time
	for three representative values of  $\theta$, including $\theta_c$, and $\rho=1/\sqrt{3}$, which is the case of the standard Grover walk \cite{inui2005}. It clearly depicts that
	when $\theta=\theta_c$
	the wavepacket tends to a delocalized regime, much like the two-state Hadamard quantum walk. As soon as $\ket{\sigma^+}$
	is taken into account ($\theta \neq \theta_c$) a finite fraction of the wavepacket remains trapped in the origin.  
	
	
	
	\begin{figure}[t!]
		\centering
		\includegraphics[width = 0.45 \textwidth]{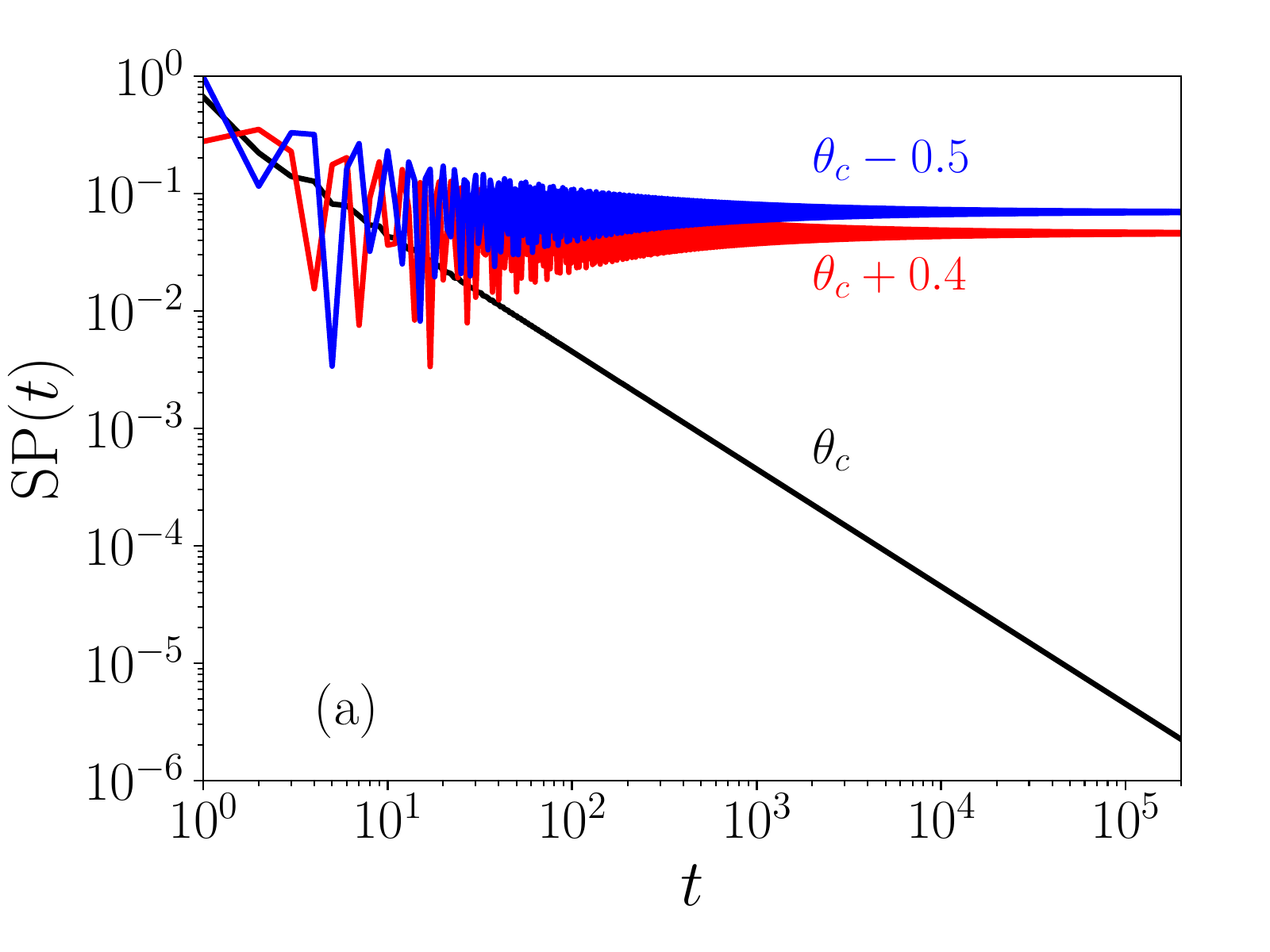}
		\includegraphics[width = 0.45 \textwidth]{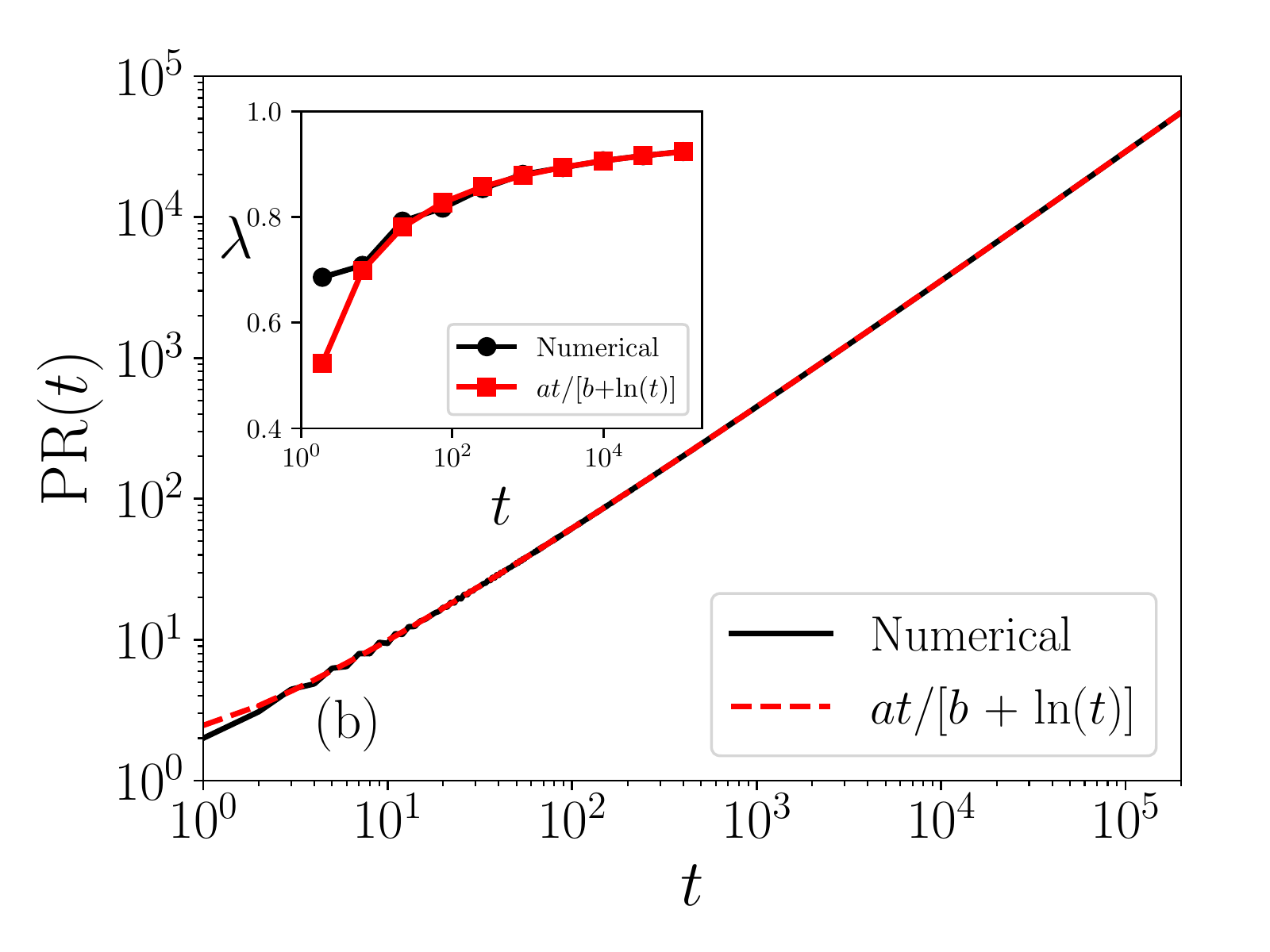}
		\caption{(a) Time evolution of the survival probability (SP) for three mixing angles and $\rho = 1/\sqrt{3}$. In the absence of localization ($\theta_c$), SP decays as $1/t$. (b) The participation ratio (PR) at $\theta_c$. It grows according to the analytical prediction $\text{PR}(\theta_c)=at/[b + \text{ln}(t)]$, with $a = 3.798$ and $b = 1.54$. The inset shows the time-dependence of the effective power-law exponent, evidencing the slow convergence to the linear behavior}
		\label{Fig.2}			
	\end{figure}

	To better characterize the dynamical behavior in the vicinity of $\theta_c$, we shall take a look 
	at the survival probability $\mathrm{SP}(t) \equiv P_{x=0}(t)$ in Fig. \ref{Fig.2}(a).
	In the long-time regime, the survival probability saturates at a finite value due to onset of localization while $\text{SP} \sim t^{-1}$
	when the mixing angle is right at $\theta_c$, as expected. This power-law decay in time of the survival probability is consistent with Eq. (\ref{Eq.3}). 
	
	Another powerful tool to 
	portray the walker dynamics is the participation ratio
	\begin{equation}
		\text{PR}(t) = \frac{1}{\sum_{x}  P_x(t)^{2}},
		\label{Eq.6}
	\end{equation}
	which accounts for the fraction of the wavepacket spreading along the lattice.
	Whenever the system has a localized component, the participation ratio  converges to a finite value, since this term will take over the sum in Eq. (\ref{Eq.6}). In the absence of that, it will become of the order of the spacial extension of the spreading wavepacket and, as such, continuously grow in time.
	Since the three-state quantum walk has an intrinsic localization in the initial site for all $\theta\neq \theta_c$, one expects $\text{PR}$ to saturate in the long-time regime. When the initial coin state is exactly $|\sigma_1^{-}\rangle$, the long-time behavior of the participation ratio can be extracted from Eq. (\ref{Eq.3}). Assuming that the spacial extension of the wavepacket becomes too large for long times, the discrete sum required to compute $\text{PR}$ can be replaced by a continuous integral, with
	the asymptotic spacial distribution $\omega({\nu})/t$. Integration by partial fractions then gives (see details in the Appendix)
	\begin{equation}
		\text{PR}(t) = \frac{at}{b + \text{ln}(t)},
		\label{Eq.7}
	\end{equation}
	where $a=a(\rho)$ and $b=b(\rho)$ are constants. This linear growth with logarithmic correction is illustrated in Fig. \ref{Fig.2}(b) for the Grover walk.

	\begin{figure}[t!]
		\centering
		\includegraphics[width = 0.45 \textwidth]{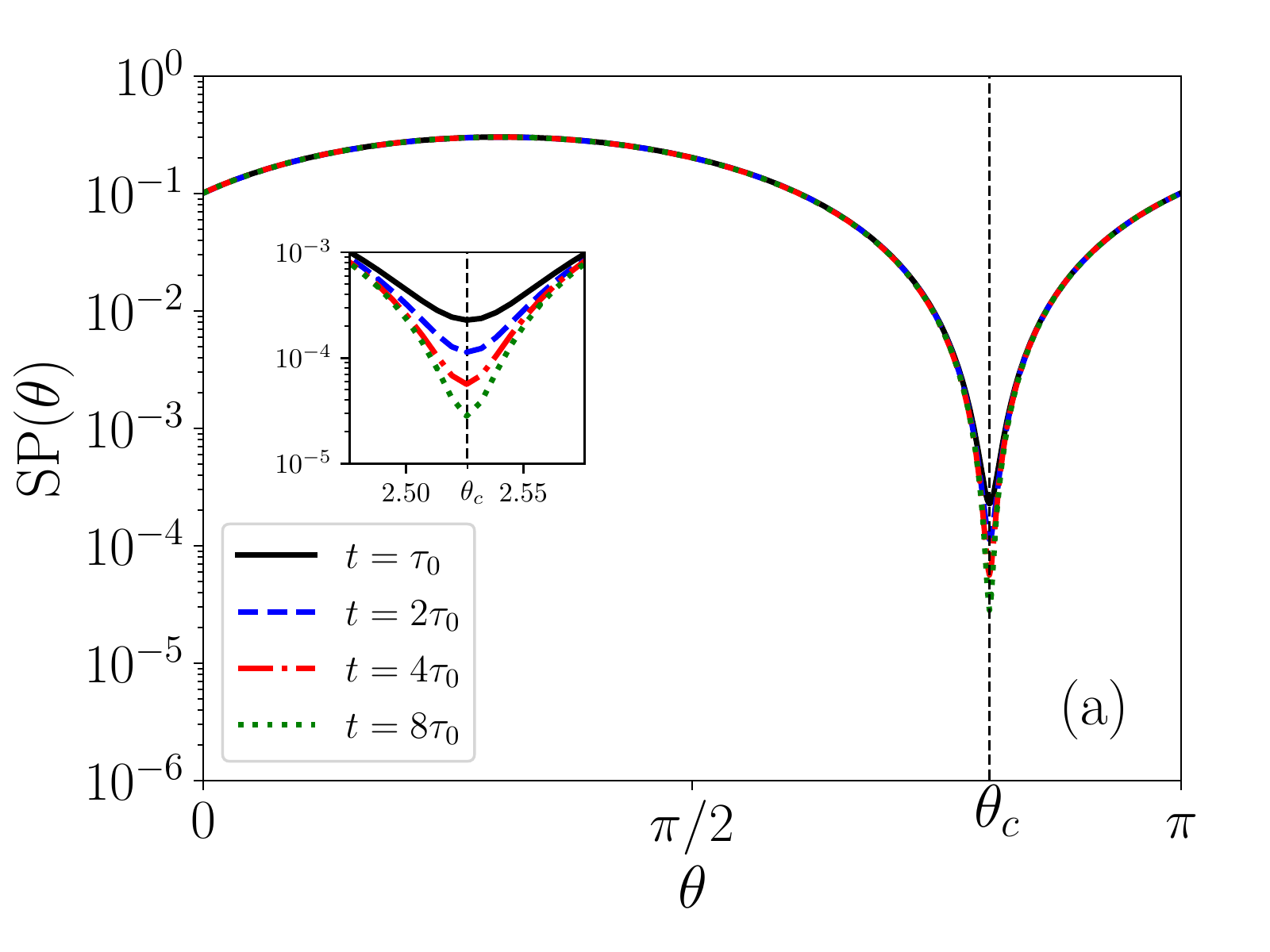}
		\includegraphics[width = 0.45 \textwidth]{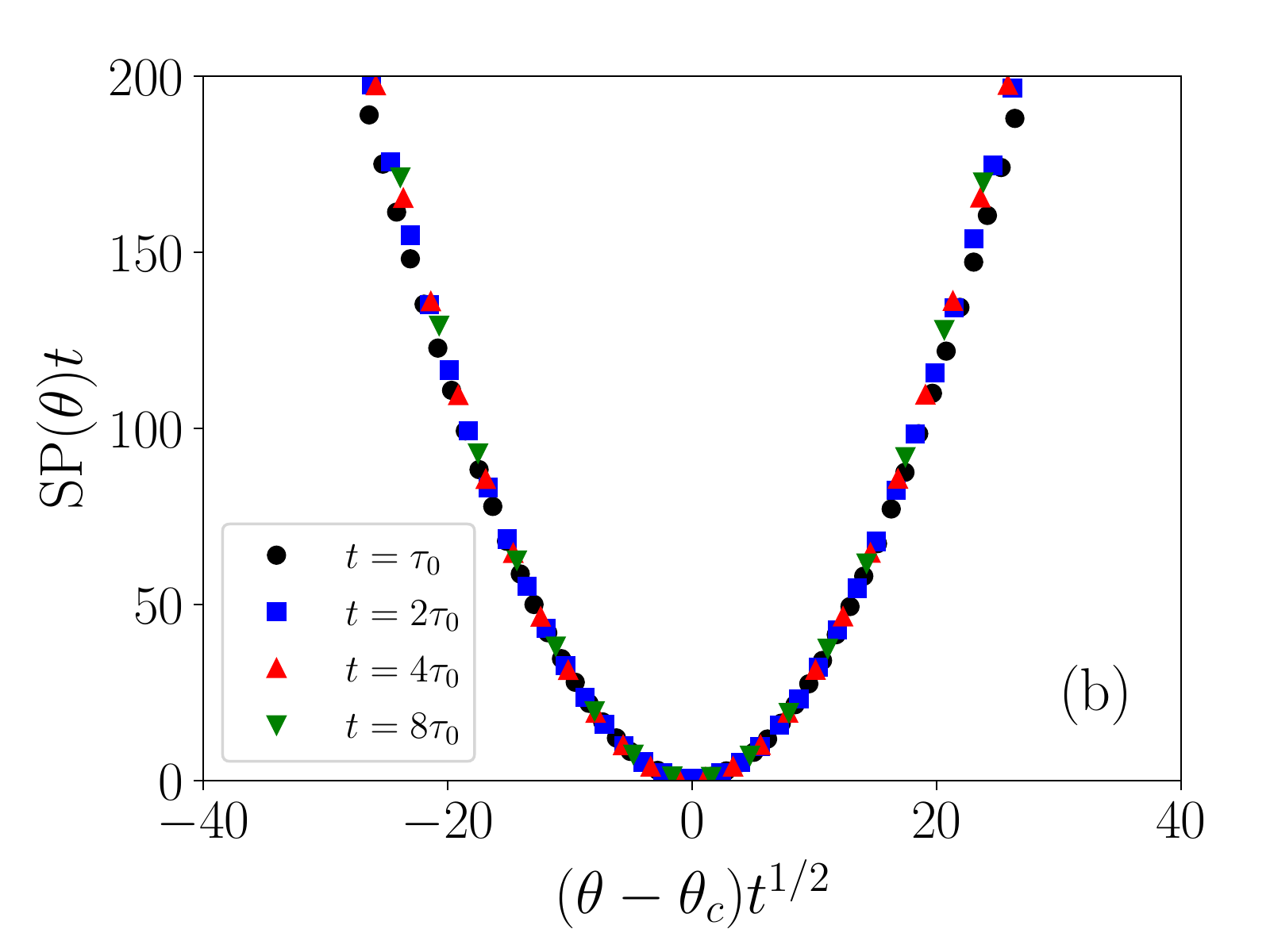}
		\caption{(a) Survival probability as a function of $\theta$ evaluated at distinct time steps, with $\tau_0 = 2000$.
			It converges quickly to a finite value, except at the characteristic angle $\theta_c$. (b) Collapse of $\text{SP}$ in the vicinity of $\theta_c$, showing a universal scaling law.}
		\label{Fig.3}
	\end{figure}

	\begin{figure}[t!]
		\centering
		\includegraphics[width = 0.45 \textwidth]{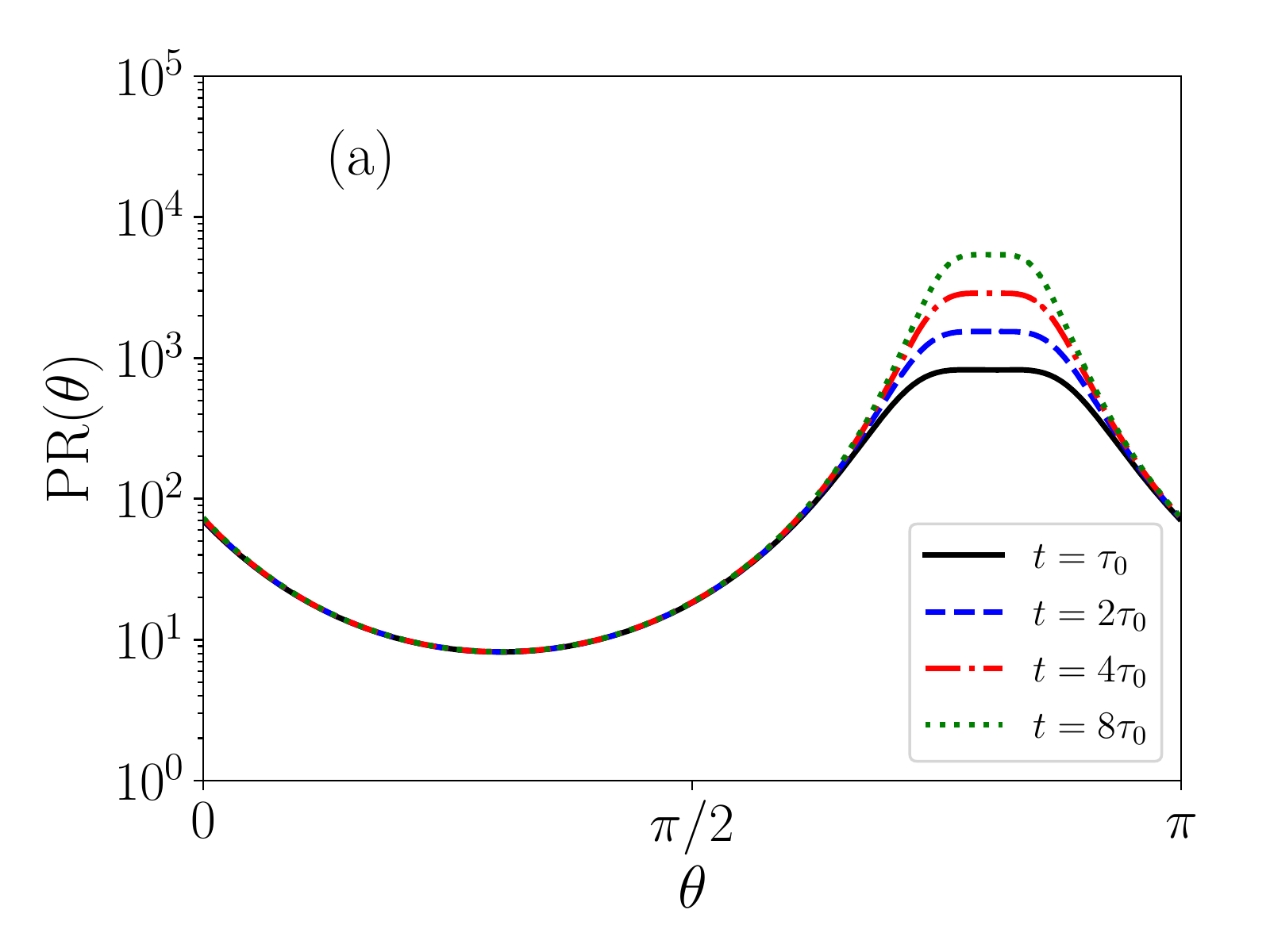}
		\includegraphics[width = 0.45 \textwidth]{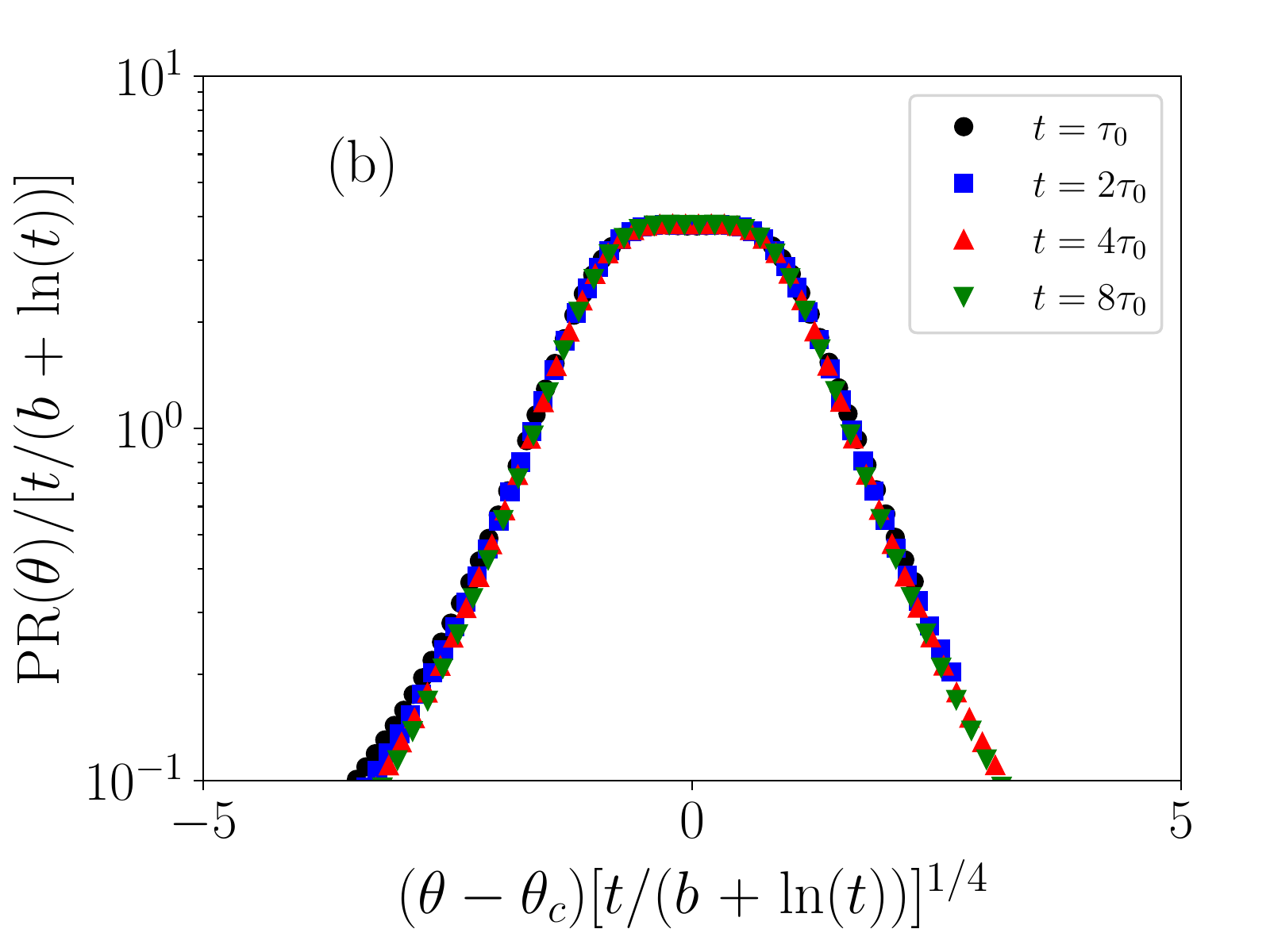}
		\caption{(a) Participation ratio as a function of $\theta$ for distinct time steps, with $\tau_0 = 2000$. (b) Data collapse unveiling a universal scaling law in the vicinity of $\theta_c$.}
		\label{Fig.4}
	\end{figure}
	
	{It is important to stress here that the presence of such logarithmic correction has to be taken into account carefully as the convergence to the ultimate linear power-law growth is very slow. Numerically, the effective power-law behavior  is estimated by the slope of the $\ln{\mathrm{PR}} \times \ln{t}$ curve in a given time interval. As the presence of a logarithmic correction implies in a slowly varying effective exponent it is useful to write $\mathrm{PR} \propto t^{\lambda(t)}$, with
		\begin{equation}
			\lambda(t)=\frac{d\ln{\mathrm{PR}}}{d\ln{t}},
		\end{equation} 
		in which a term proportional to the rate of variation of $\lambda(t) $ was disregarded assuming $d(\ln{\lambda})/d(\ln{t}) << \ln{\lambda}/\ln{t}$. The asymptotic behavior given by Eq. (\ref{Eq.7}) implies that $\lambda(t) = 1 - 1/(b+\ln{t})$, evidencing the very slow logarithmic convergence $\lambda(t\rightarrow\infty)\rightarrow 1$. }
	
	%
	%
	In the inset of Fig. \ref{Fig.2}(b) we show the time evolution of the $\lambda$ computed via numerical calculation of $\text{PR}$ alongside the one derived from Eq. (\ref{Eq.7}). 
	Note that the asymptotic convergence to $\lambda=1$ is actually very slow. As a matter of fact, such logarithmic correction on the time-evolution of the participation ratio is quite common in quantum walks with no localized component (e.g. two-state quantum walks)
	featuring group-velocity distribution singularities 
	such as those present in Eq. (\ref{Eq.3}).
	Furthermore, we would also like highlight that the sublinear time evolution of the participation ratio reported in previous studies \cite{omanakuttan2018} is nothing more than an effective description limited to a specific time interval.
	It hinders the logarithmic
	correction unveiled here, which is the true asymptotic behavior. 
	
	Let us now turn our attention to the dynamics occurring at the vicinity of $\theta_c$.
	In Fig. \ref{Fig.3}(a), we show the survival probability of the Grover walk at distinct times as a function of $\theta$ so as to cover a variety of initial states. 
	Once again, SP saturates to a fixed value except at the vicinity of the critical parameter $\theta_c$, that is where it scales as $\text{SP} \sim t^{-1}$.     
	In this region data can be collapsed into a single universal curve by relying on the fact that in the long time limit, SP decays as $\text{SP}\propto (\theta-\theta_c)^2$. This quadratic behavior takes place as the overlap between $|\sigma^+\rangle$ 
	and the initial state $\alpha$ vanishes linearly.
	In association with the $1/t$ decay at $\theta_c$ discussed before, we can thus rewrite the survival probability in a single-parameter scaling form as
	\begin{equation}
		\text{SP}(\theta , t) = t^{-1}f[(\theta-\theta_c)t^{1/2}],   
	\end{equation}
	with $f(0)$ being a constant and $f(\eta \gg 1)\propto \eta^2$. 
	This is depicted in \ref{Fig.3}(b), where we plot $t\times \text{SP}$ as a function of the scaling variable $(\theta-\theta_c)t^{1/2}$. The
	collapse resulting from the curves
	obtained from distinct time steps corroborates the proposed single parameter scaling in the vicinity of $\theta_c$.  
	Most importantly, 
	this is independent of the coin parameter $\rho$ and thus can be considered as a universal dynamical scaling law for the present  three-state quantum walk model.
	
	We also analyze the $\text{PR} = \text{PR}(\theta)$ for distinct time steps as illustrated in Fig. \ref{Fig.4}(a). As discussed previously, $\text{PR}$ saturates to a finite value due to the localization occurring at the origin. It is only at $\theta_c$ that PR grows linearly with the reported logarithmic correction. However, finite-time corrections take place  close to $\theta_c$. In the long-time regime $t\rightarrow\infty$, $\text{PR}$ shall diverge as $\text{PR}\propto (\theta-\theta_c)^{-4}$. Such law results from the very definition of $\text{PR}$ that takes the squared survival probability as input. The participation ratio can also be written in a single parameter scaling form that incorporates both asymptotic behaviors, such as
	\begin{equation}
		\text{PR}(\theta , t) = \tilde{t}g[(\theta-\theta_c){\tilde t}^{1/4}],
	\end{equation}
	where $\tilde{t}=t/(b+\ln{t})$, $g(0)$ is a constant, and $g(\eta \gg 1)\propto \eta^{-4}$. 
	Data collapse is shown in Fig. \ref{Fig.4}(b), where we plot $\text{PR}/\tilde{t}$ as a function of the scaling variable $(\theta-\theta_c)\tilde{t}^{1/4}$. Once again, the collapse is accurate thus supporting both the single-parameter dynamic scaling hypothesis and the logarithmic correction to the relevant time scale.
	
\begin{figure}[t!]
		\centering
		\includegraphics[width =\linewidth]{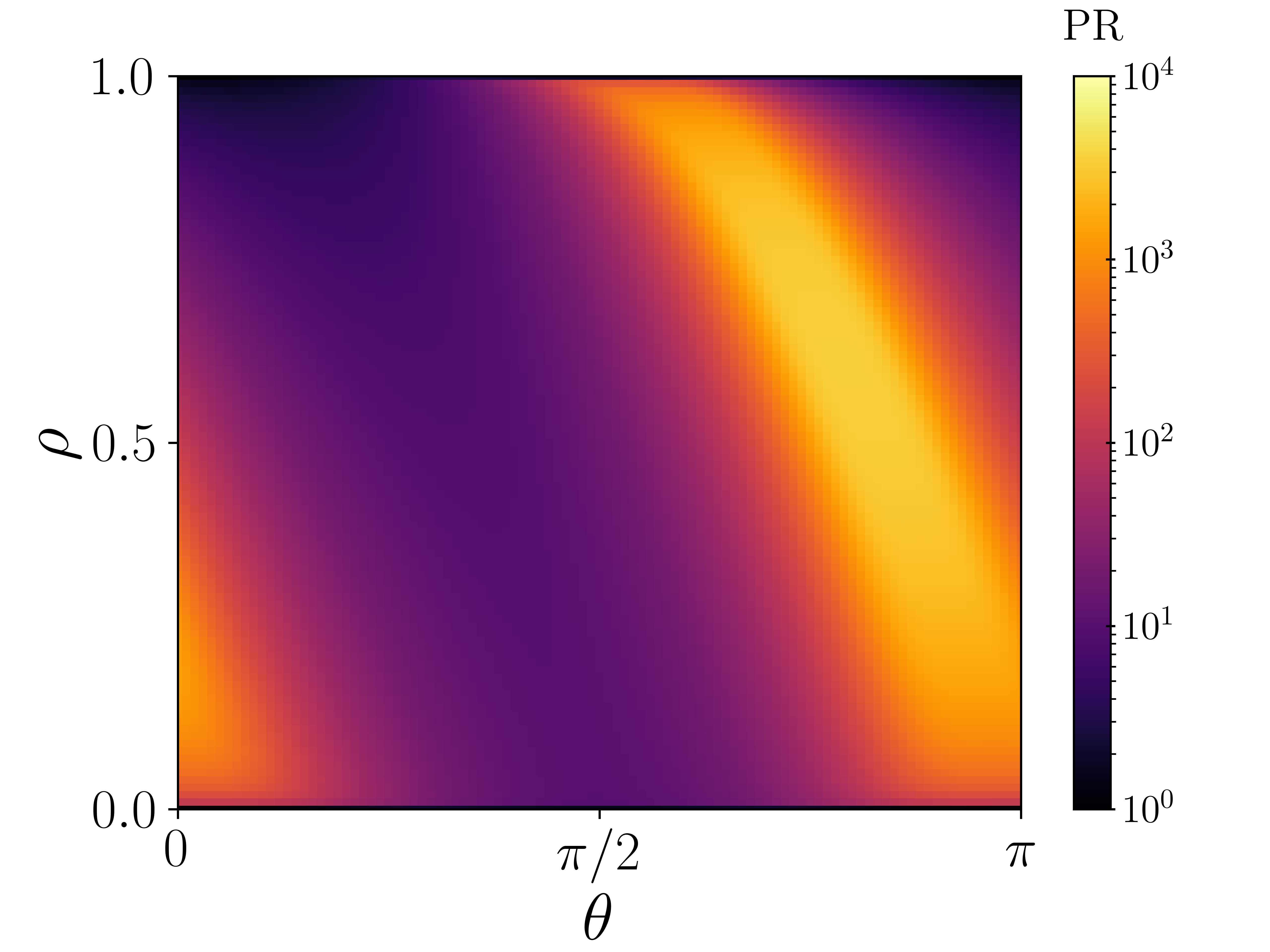}
		\caption{Density plot of the participation ratio $\text{PR}$ in the $\rho$ vs $\theta$ parameter space evaluated at $t=10^4$ (long-time regime). Delocalization is pronounced at $\theta_c=\cos^{-1}(-{\sqrt{1-\rho^2}})$, with maximum wavepacket spreading being reached as the Grover walk condition $\rho=1/\sqrt{3}$ is approached. }
		\label{Fig.5}
	\end{figure}

	Last but not least, we report the dependence of the wavepacket spreading dynamics on the coin parameter $\rho$. A density plot of the participation ratio as a function of both $\rho$ and $\theta$ evaluated at the long-time regime is shown in Fig. \ref{Fig.5}. It becomes large at the locus $\theta_c=\cos^{-1}(-{\sqrt{1-\rho^2}})$. However, the degree of wavepacket delocalization in nonuniform and reaches its maximum when the coin parameter approaches that of the Grover walk, $\rho=1/\sqrt{3}$.



	\section{Concluding remarks}
	
	We have studied the dynamics of a general three-state quantum walk, where the coin operator allows for a continuous tunning from a non-propagating regime $\rho=0$ to a non-mixing regime $\rho=1$ between the left and right side components. 
	We set as input an effective two-level state whose relative phase 
	parametrized by an angle $\theta$ weights the contribution of the coin eigenstates that leads to delocalized and localized dynamical regimes.

	Full delocalization was shown to occur only at $\theta = \theta_c$, where the survival probability $\text{SP}\propto 1/t$.
	A detailed analysis of the wavepacket dynamics was then carried out
	around that point and we unveiled a single-parameter dynamical scaling law of the form $\text{SP} = t^{-1}f(\eta)$, with $\eta=(\theta-\theta_c)t^{1/2}$. The scaling function $f(\eta \gg 1)\propto \eta^2$ accounts for the long-time quadratic decay of the survival probability as $\theta_c$ is approached.
	
	We also showed that the participation ratio $\text{PR}$ 
	grows linearly in time with a logarithmic correction of the form $\text{PR}\propto t/(b+\ln{t})$.
	This feature is present
	in delocalized quantum walks, such as the two-state Hadamard quantum walk, having an asymptotic distribution
	with group-velocity distribution singularities as found in Eq. (\ref{Eq.3}). 
	This logarithmic correction
	uncovers the actual mechanism underneath
	the effective sublinear dynamics reported in \cite{omanakuttan2018}, for instance. 
	
	The participation ratio was found to satisfy a single-parameter scaling law $\text{PR}\propto \tilde{t}g(\eta)$ with $\tilde t = t/(b+\ln{t})$ and $\eta=(\theta-\theta_c)\tilde{t}^{1/4}$. The scaling function $g(\eta \gg 1) \propto \eta^4$, thus accounting for the long-time divergence of $\text{PR}\propto (\theta-\theta_c)^{ -4}$ in the vicinity of $\theta_c$. 
	
	The above scaling laws worked out here are universal and hold for all values of the coin parameter $\rho$ for the three-state quantum walk. However, the degree of delocalization at $\theta_c$ depends nonmonotonicaly on $\rho$, reaching the maximum as the Grover walk condition $\rho=1/\sqrt{3}$ is fulfilled. 
	The finite-time scaling analysis 
	employed here 
	can be extended to bring about
	hidden features underlying several families of quantum walks, including those whose dynamics is already well established. This is of great importance for the design of quantum algorithms. 
	Seeking for universal dynamical laws in discrete-time quantum walks 
	also make for a better understanding 
	over their connections with Hamiltonian models, allowing for progress in physical implementations. 

	\section{Acknowledgments}
	This work was partially supported by CAPES (Coordena\c{c}\~ ao de Aperfei\c{c}oamento de Pessoal do N\'{\i}vel Superior), CNPq (Conselho Nacional de Desenvolvimento Cient\'{\i}fico e
	Tecnol\'ogico), and FAPEAL (Funda\c{c}\~ ao de Apoio \`a Pesquisa do
	Estado de Alagoas).

	\section*{Appendix}
	
	\begin{figure}[t!]
		\centering
		\includegraphics[width=0.49 \textwidth]{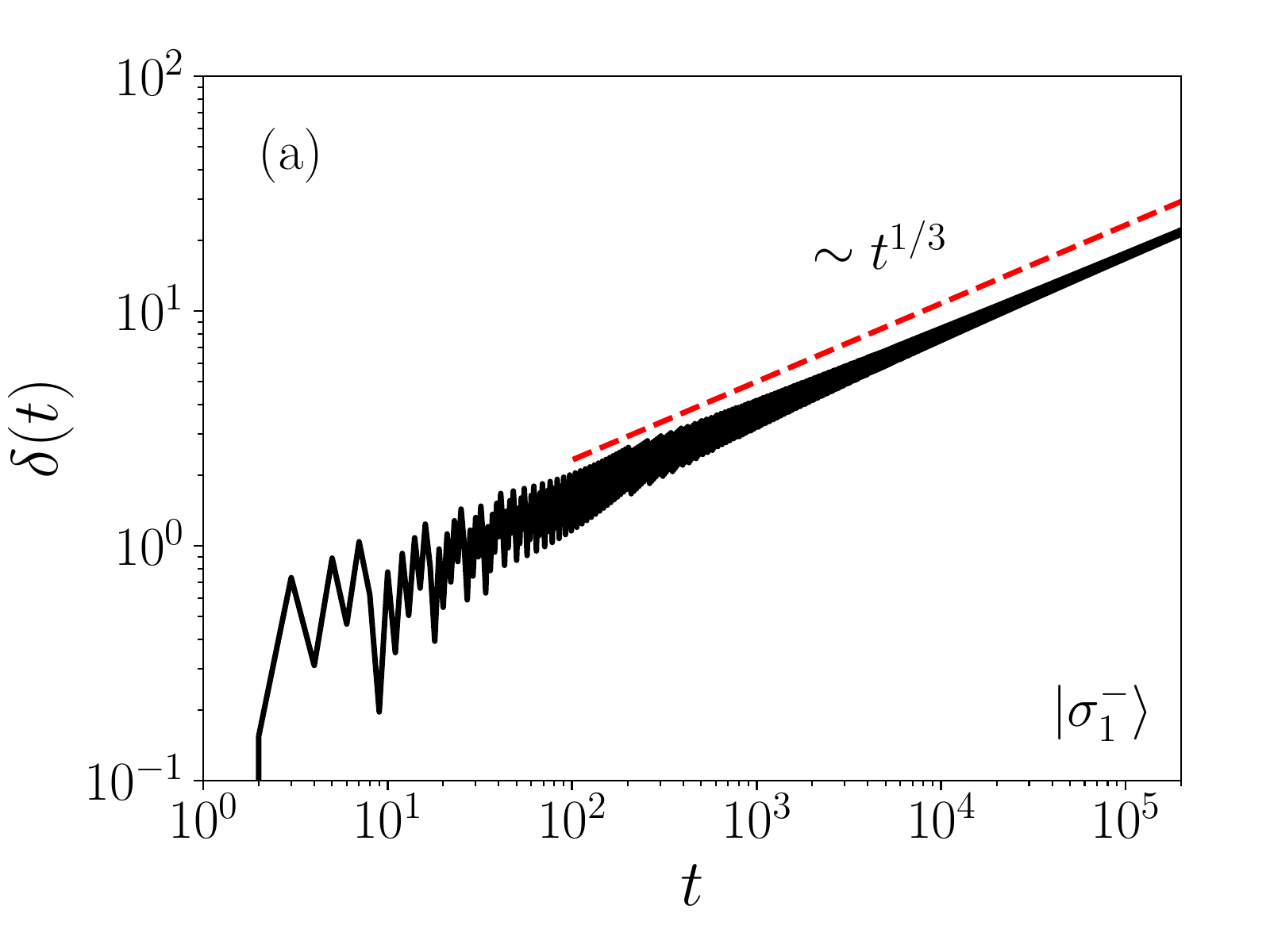}
		\includegraphics[width=0.49 \textwidth]{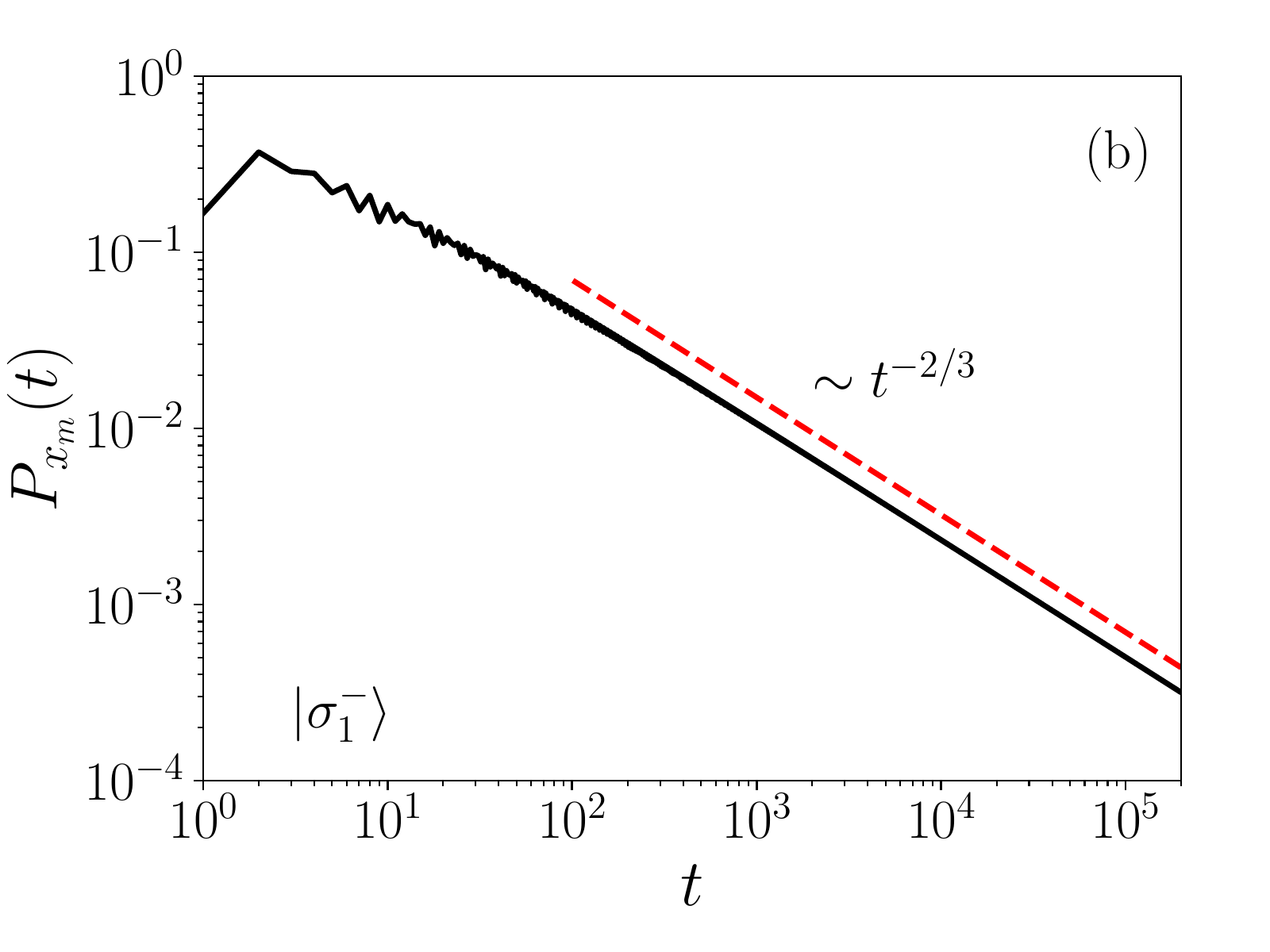}
		\caption{(a) Sublinear time-evolution of $\delta(t)=\rho t-x_m$, evidencing $\delta \propto t^{1/3}$; (b) Time-dependence of the occupation probability at the wave-front $P_{x_m}$ indicating its power-law decay $P_{x_m}\propto t^{-2/3}$.}
		\label{Fig.6}
	\end{figure}

	In this appendix, we carry out the derivation of Eq. (\ref{Eq.7}).
	Considering the initial state $|\sigma_1^-\rangle$, the long-time behavior of the participation ratio  can be analytically extracted using the group-velocity density given by Eq. (\ref{Eq.3}). The inverse participation ratio is given by
	\begin{equation} \label{eqap}
		[\mathrm{PR}(t)]^{-1} = \sum_x P_x(t)^2 = \sum_x[\omega(\nu)/t]^2 ,
	\end{equation}
	with $\nu=x/t$ and the sum extending over the wave-packet support whose fronts evolve ballistically in time with velocity $\rho$.  The wave-front positions are at $x_m=\pm(\rho t-\delta)$, with $\delta$ being a positive sublinear correction. In Fig. \ref{Fig.6}(a) we depict the numerically obtained time-dependence  $\delta(t)\propto t^{1/3}$ for which the wave-front was considered to be the position of the wave distribution maximum. Such sublinear correction actually avoids a true divergence of the probability density at the wave-front. According to Eq. (\ref{Eq.3}), $\omega(\nu_m=x_m/t)\propto\sqrt{t/\delta(t)}$ in the long-time regime. This implies that the probability density at the wave-front is $P_{x{_m}}(t)\propto t^{-2/3}$ as made clear in Fig. \ref{Fig.6}(b). Replacing the sum over $x$ with a sum over $\nu$ in Eq. \ref{eqap}, we get
	\begin{equation}
		[\mathrm{PR}(t)]^{-1} = \sum_{\nu=-(\rho-\delta/t)}^{\rho-\delta/t}[\omega(\nu)/t]^2 ,
	\end{equation}
	where the interval between successive values of $\nu$ is $\Delta\nu=1/t$. Therefore, in the long-time regime, the last sum can be turned into an integral of the form
	\begin{eqnarray}
		[\mathrm{PR}(t)]^{-1} &=& \int_{-(\rho-\delta/t)}^{\rho-\delta/t} \frac{\omega^2}{t} d\nu \nonumber \\ ~&=& \frac{1-\rho^2}{\pi^2 t} \int_{-(\rho-\delta/t)}^{\rho-\delta/t} \frac{d\nu}{(1-\nu^2)^2(\rho^2-\nu^2)} .
	\end{eqnarray}
	One can then highlight its time-dependence via the partial fraction decomposition
	\begin{equation}
		\frac{1}{(1-\nu^2)^2(\rho^2-\nu^2)} = \frac{A}{(1-\nu^2)^2}+\frac{B}{1-\nu^2}+\frac{C}{\rho-\nu}+\frac{D}{\rho+\nu}.
	\end{equation}
	The integrals of the first two terms converge to finite values as $t\rightarrow\infty$ while the last two diverge as $\ln{t}$ due to the group-velocity distribution singularities at $\nu=\pm\rho$.  Consequently, the long-time  inverse participation ratio decays as $[\mathrm{PR}(t)]^{-1} = (1/at)(b + \ln{t})$, unveling a logarithmic correction to the linear long-time evolution of the participation ratio.


\begin{thebibliography}{}
		
		\bibitem{aharonov1993}
		Y. Aharonov,  L. Davidovich and N. Zagury, Phys. Rev. A \textbf{48}, 1687 (1993).
		
		\bibitem{shenvi2003}
		N. Shenvi, J. Kempe and K. Birgitta Whaley,  Phys. Rev. A \textbf{67}, 052307 (2003).
		
		\bibitem{childs2004}
		A. M. Childs and J. Goldstone,  Phys. Rev. A \textbf{70}, 022314 (2004).
		
		\bibitem{childs2014}
		A. M. Childs and  Y. Ge, Phys. Rev. A, \textbf{89}, 052337 (2014).
		
		\bibitem{wong2015}
		T. G. Wong, J. Phys. A: Math. Theor. \textbf{48}, 435304 (2015).		
		
		\bibitem{wong2018}
		T. G. Wong, Quantum Inf. Process. \textbf{17}, 68 (2018).
		
		\bibitem{zhan2014}
		X. Zhan, H. Qin, Z. H. Bian, J. Li, and P. Xue, Phys. Rev. A \textbf{90},
		012331 (2014).
		
		\bibitem{stefanak2016}
		M. \v{S}tefa\v{n}\'ak and S. Skoup\'y, Phys. Rev. A \textbf{94}, 022301 (2016).
		
		\bibitem{kurzynsky2011}
		P. Kurzy\'nski and A. W\'ojcik, Phys. Rev. A \textbf{83}, 062315 (2011).
		
		
		\bibitem{knight2003}
		P. L. Knight, E. Rold\'an, and J. E. Sipe, Phys. Rev. A \textbf{68},
		020301(R) (2003).
		
		
		\bibitem{childs2009}
		A. M. Childs, Phys. Rev. Lett. \textbf{102}, 180501 (2009).
		
		\bibitem{lovett2010}
		N. B. Lovett, S. Cooper, M. Everitt, M. Trevers, and V. Kendon, Phys. Rev. A \textbf{81}, 042330 (2010).
		
		\bibitem{kempe2003} J. Kempe, Contemp. Phys. \textbf{44}, 307 (2003).
		
		\bibitem{venegas2012}  S. E. Venegas-Andraca, Quantum Inf. Process. \textbf{11}, 1015 (2012).
		
		
		\bibitem{vakulchyk2019} I. Vakulchyk, M. V. Fistul, and S. Flach, Phys. Rev. Lett. \textbf{122}, 040501 (2019).
		
		\bibitem{chandrashekar2008}
		C. M. Chandrashekar and R. Laflamme, Phys. Rev. A \textbf{78},
		022314 (2008).
		
		\bibitem{wang2019}
		K. Wang, X. Qiu, L. Xiao, X. Zhan, Z. Bian, W. Yi, and P. Xue, Phys. Rev. Lett. \textbf{122}, 020501 (2019).
		
		\bibitem{xwang2018}
		X. P. Wang, L. Xiao, X. Z. Qiu, K. K. Wang, W. Yi, and
		P. Xue, Phys. Rev. A \textbf{98}, 013835 (2018).
		
		\bibitem{navarrete2007}
		C. Navarrete-Benlloch, A. P\'erez, and E. Rold\'an, Phys. Rev. A
		\textbf{75}, 062333 (2007)
		
		\bibitem{buarque2020}
		A. R. C. Buarque and W. S. Dias, Phys. Rev. A \textbf{101}, 023802
		(2020).
		
		\bibitem{mendonca2020}
		J. P. Mendon\c{c}a, F. A. B. F. de Moura, M. L. Lyra, and G. M. A.
		Almeida, Phys. Rev. A \textbf{101}, 062335 (2020).
		
		\bibitem{buarque2021} A. R. C. Buarque and W. S. Dias, Phys. Rev. A \textbf{103}, 042213 (2021).
		
		
		\bibitem{kitagawa2010}
		T. Kitagawa, M. S. Rudner, E. Berg, and E. Demler, Phys. Rev.
		A \textbf{82}, 033429 (2010)
		
		\bibitem{asboth2012}
		J. K. Asb\'oth, Phys. Rev. B \textbf{86}, 195414 (2012).
		
		\bibitem{rakovszky2015}
		T. Rakovszky and J. K. Asb\'oth, Phys. Rev. A \textbf{92}, 052311 (2015)
		
		\bibitem{obuse2011} H.   Obuse   and   N.   Kawakami,   Phys.   Rev.   B   \textbf{84}, 195139  (2011).
		
		\bibitem{mendes2019} C. V. C. Mendes, G. M. A. Almeida, M. L. Lyra, and F. A. B. F. de Moura, Phys. Rev. E \textbf{99}, 022117 (2019).
		
		\bibitem{buarque2019} A. R. C. Buarque and W. S. Dias, Phys. Rev. E \textbf{100}, 032106 (2019).
		
		\bibitem{queiros2020} M. A. Pires and S. M. Duarte Queir\'os, Phys. Rev. E \textbf{102}, 012104 (2020).
		
		\bibitem{inui2005}
		N. Inui, N. Konno, and E. Segawa, Phys. Rev. E \textbf{72}, 056112
		(2005).
		
		\bibitem{falkner2014}
		S. Falkner and S. Boettcher,  Phys. Rev. A \textbf{90}, 012307 (2014).
		
		\bibitem{konno2014}
		N. Konno,  Quantum Inf. Process. \textbf{13}, 1103-1125 (2014).
		
		
		\bibitem{stefanak2012}
		M. \v{S}tefa\v{n}\'ak, I. Bezd\v{e}kov\'a and I. Jex, Eur. Phys. J. D \textbf{66},
		142 (2012).
		
		\bibitem{stefanak2014}
		M. \v{S}tefa\v{n}\'ak, I. Bezd\v{e}kov\'a and I. Jex, Phys. Rev. A \textbf{90}, 012342 (2014).
		
		\bibitem{wong2017}
		T. G. Wong, J. Phys. A: Math. Theor. \textbf{50}, 475301 (2017).
		
		\bibitem{saha2021}
		A. Saha, S. B. Mandal, D. Saha and A.  Chakrabarti, IEEE Trans. Quant. Eng. \textbf{2}, 3102012 (2021).
		
		
		\bibitem{omanakuttan2018}
		S. Omanakuttan and A. Lakshminarayan, J. Phys. A: Math.
		Theor. \textbf{51}, 385306 (2018).
		
		\bibitem{grimmet2004} G. Grimmett, S. Janson, and P. F. Scudo, Phys. Rev. E \textbf{69}, 026119 (2004). 
		
		\bibitem{machida2015}
		T. Machida, Quantum Information \& Computation \textbf{15}, 406 (2015).
		
		
		
	\end{thebibliography}
\end{document}